\documentclass[twocolumn,amsmath,amssymb]{revtex4}
\usepackage{graphicx}
\usepackage{dcolumn}
\usepackage{bm}
\begin{document}

\title{Excitation Gap of Fractal Quantum Hall States in Graphene}
\author{Wenchen Luo}
\affiliation{Departement of Physics and Astronomy, University of Manitoba, Winnipeg,
Canada R3T 2N2}
\author{Tapash Chakraborty }
\affiliation{Departement of Physics and Astronomy, University of Manitoba, Winnipeg,
Canada R3T 2N2}

\begin{abstract}
In the presence of a magnetic field and an external periodic potential, the
Landau level spectrum of a two-dimensional electron gas exhibits a fractal
pattern in the energy spectrum which is described as the Hofstadter's
butterfly. In this work, we develop a Hartree-Fock theory to deal with the
electron-electron interaction in the Hofstadter's butterfly state in a
finite-size graphene with periodic boundary conditions, in which we include
both spin and valley degrees of freedom. We then treat the butterfly state
as an electron crystal so that we could obtain the order parameters of
the crystal in the momentum space and also in an infinite sample. The
excitation gaps obtained in the infinite sample is comparable to those in
the finite-size study, and agree with a recent experimental observation.
\end{abstract}

\date{\today }
\maketitle

In a strong perpendicular magnetic field, the energy spectrum of a
two-dimensional electron gas (2DEG) splits into a series of Landau levels.
If a periodic potential is also added to this system, the noninteracting
energy spectrum then displays an intricate fractal pattern known as the
Hoftstadter's butterfly \cite{hofstadter}. In the quantum Hall effect
regime, the Coulomb interaction between electrons plays an important role
both in the ground state and in the low-energy excitations. Theoretically,
the Coulomb interaction effects on the butterfly states were investigated in
the Hartree or mean field approximation \cite{vidar,doh,vadym}, or with the
electron correlations \cite{pfannkuche,areg}. In the Hartree approximation,
the electrons are classical particles with negative charge and repel each
other. This approximation is unable to deal with a spin/valley system such
as graphene if the spin is not polarized, since the exchange interaction is
not taken into consideration. In the exact diagonalization scheme, the
electron correlations are completely included, and hence more adaptable to
the fractional quantum Hall effect \cite{fqhe}, but then only the
finite-size systems can be handled in this scheme. Here we consider the
Hartree-Fock approximation (HFA) to deal with the Coulomb interaction and
the spin/valley system in graphene. Further, we work in the momentum space
to study an infinte system.

After a moir\'{e} pattern is fabricated by two misaligned honeycomb
lattices by graphene and the boron nitride (BN) substate 
\cite{xue,decker,yankowitz,ponomarenko}, the fractal band structure and the
transport properties of the butterfly states were finally revealed in recent
experiments \cite{dean,yu}. Theoretical studies of the butterfly states in
monolayer graphene and bilayer graphene have exhibited very rich physics 
\cite{vadym,areg,nemec,bistritzer,moon,godfrey,diez}.

In contrast to the conventional 2DEG in a semiconductor, electrons in
graphene must be described by the Dirac equation with an extra valley degree
of freedom \cite{book,abergel,rmp}. In this case, the Hartree approximation
may not be enough to describe the behavior of the electrons if we consider
both the spin and the valley. Indeed, the excitation gap measured in a
recent experiment \cite{yu} can not be explained either in the
noninteracting picture or in the Hartree approximation. Here, we
develop a Hartree-Fock approximation to calculate the energy gaps for
integer filling factors. We also employ a method involving the
crystalline state to explain the experimental results.

The Hofstadter states are not affected much by the geometry of the external
potential. For simplicity and without loss of generality we introduce in the
graphene Hamiltonian a square periodic potential \cite{tapash},
$V_{\rm ext}\left(x,y\right)=V_{0}\left[\cos\left(q_{0}x\right) + \cos\left( 
q_{0}y\right) \right],$ where $V_{0}$ is the amplitude of the potential and 
$q _{0}=2\pi /a _{0}$ with a period of the potential $a _{0}$. The
Hamiltonian of the 2DEG in such a system is then given by $H=H_{0} + V_{C}
+ V_{\mathrm{ext}}.$ The noninteracting Hamiltonian \cite{book,abergel,rmp,goerbig} 
without the external potential is 
\begin{equation}
H_{0}=v_{F}\left( 
\begin{array}{cc}
0 & P_{x}-i\eta P_{y} \\ 
P_{x}+i\eta P_{y} & 0
\end{array}
\right),
\end{equation}
where $\eta=\pm1$ for $K$ and $K^\prime$ valley respectively, $v_{F}$ is the 
Fermi velocity, and $\mathbf{P}=\mathbf{p}+e\mathbf{A}$ is
the canonical momentum. We choose the Landau gauge to define the vector
potential $\mathbf{A}=\left( 0,Bx\right) $, and $V_{C}$ is the
electron-electron interaction.

The period of the external potential is large enough experimentally \cite{dean,yu} 
(larger than $10$nm), so that the valley mixing can be neglected.
The energy bandwidth is narrow when $V_{0}$ is not large. The Landau level
(LL) mixing is also very weak and is not considered here. We first
diagonalize the noninteracting Hamiltonian with the external potential, 
$\widetilde{H}_{0}=H_{0}+V_{\mathrm{ext}}$ by using the basis $\left\{ \phi
_{X}^{\sigma }\right\} $, where $\phi _{X}^{\sigma }$ is the wave function
of the $n$-th LL with valley-spin index $\sigma $ and the guiding center $X$ 
\cite{book,abergel,rmp,goerbig}. Then we could obtain $4N_{\phi }$ eigenvectors: 
\begin{equation}
\left( 
\begin{array}{ccccc}
c_{i,X_{1}}^{\sigma _{1}}\phi _{X_{1}}^{\sigma _{1}} & \ldots & 
c_{i,X_{N_{\phi }}}^{\sigma _{1}}\phi _{X_{N_{\phi }}}^{\sigma _{1}} & \ldots
& c_{i,X_{N_{\phi }}}^{\sigma _{4}}\phi _{X_{N_{\phi }}}^{\sigma _{4}}
\end{array}
\right) ^{\dag },
\end{equation}
where $i=1,\ldots 4N_{\phi }$ and $4N_{\phi }$ is the degeneracy of the
Landau level in the finite sample. The coefficients $c_{i,X_{j}}^{\sigma }$
satisfy the normalization condition $\sum_{\sigma }\sum_{j=1}^{N_{\phi
}}\left\vert c_{i,X_{j}}^{\sigma }\right\vert ^{2}=1.$ Note that $i$ is also
the index of the corresponding eigenenergies with ascending order and $j$ is
the guiding center index for the $N_{\phi }$ states in a single LL.

The Hartree-Fock Coulomb interaction shall be determined self-consistently
by the coefficients $c_{i,X_j}^{\sigma }$. Let us consider the Hartree term 
$V_{H}$ and the Fock term $V_{F}$\ separately. The Hartree term 
\cite{vidar,vadym,tapash} is 
\begin{eqnarray}
&&\left\langle \sigma ^{\prime },X^{\prime }\right\vert V_{H}\left\vert
\sigma,X\right\rangle =\frac{\delta _{\sigma ,\sigma ^{\prime }}}{N_{\phi }}
\frac{e^{2}}{\kappa \ell }\sum_{i}^{\prime }\overline{\sum_{\mathbf{G}}}
\sum_{\xi }\sum_{k,l}^{N_{\phi }}\frac{1}{G\ell }  \notag \\
&&\times c_{i,k}^{\xi \ast }c_{i,l}^{\xi }M_{k,l}\left( -\mathbf{G}\right)
M_{X^{\prime },X}\left( \mathbf{G}\right) ,
\end{eqnarray}
where $\kappa $ is the dielectric constant, the summation with a prime sums
over all the states below the Fermi level, and the bar over the summation
means the term with $G=0$ is excluded. The function $M$ is defined by 
\begin{eqnarray}
M_{k,l}\left( \mathbf{G}\right) &=&\frac{\delta _{k,l+G_{y}\ell ^{2}}}{
2^{1-\delta _{n,0}}}e^{-\frac{i}{2}G_{x}\left( k+l\right) }e^{-G^{2}\ell
^{2}/4} \\
&&\times \left[ L_{\left\vert n\right\vert -1}\left( \frac{G^{2}\ell ^{2}}{2}
\right) +L_{\left\vert n\right\vert }\left( \frac{G^{2}\ell ^{2}}{2}\right) 
\right] .  \notag
\end{eqnarray}
with a Laguerre polynomial $L$ (we define $L_{n<0}=0$) and the magnetic
length $\ell =\sqrt{\hbar /\left( eB\right) }$. The Fock term is 
\begin{eqnarray}
&&\left\langle \sigma ^{\prime },X^{\prime }\right\vert V_{F}\left\vert
\sigma ,X\right\rangle =\frac{e^{2}}{\kappa \ell N_{\phi }}\sum_{i}^{\prime }
\overline{\sum_{\mathbf{G}}}\sum_{k,l}^{N_{\phi }}\frac{1}{G\ell }  \notag \\
&&\times c_{i,k}^{\sigma \ast }c_{i,l}^{\sigma ^{\prime }}M_{k,X}\left( -
\mathbf{G}\right) M_{X^{\prime },l}\left( \mathbf{G}\right) .
\end{eqnarray}
For a finite sample the momentum vectors are discrete, $\mathbf{G}=\left(
2\pi /a_{0}\right) \left( n_{x},n_{y}\right) $, where $n_{x},n_{y}$ are
integers.

Once we diagonalize the noninteracting Hamiltonian $\widetilde{H}^{}_{0}$,
we obtain a series of coefficients $c_{i,X _{j}}^{\sigma }$. For the $i$-th
eigenenergy $E _{i}$, the eigenvector is given by $c_{i,X _{j}}^{\sigma }$.
The summation $\sum^{\prime }$ in the Hartree-Fock approximation contains $
i=1\ldots N _{s}$. The filling factor is defined by $\nu =N _{s}/N _{\phi }$
. We use these coeffeicients $c_{i,X _{j}}^{\sigma }$ to compute the
Hartree-Fock Coulomb interaction. Then the full Hamiltonian $H$ is
diagonalized and a new group of the coefficients $c_{i,X _{j}}^{\sigma }$ is
obtained. By repeating this process the coefficients $c_{i,X _{j}}^{\sigma}$
and the energy spectrum can be evaluated self-consistently. The energy gap
is obtained from $\Delta=E _{N _{s}+1}-E _{N _{s}}.$ It is the gap between
the highest occupied state and the lowest unoccupied state. This gap may be
observed in transport or capacitance measurements.

The parameter $\alpha$ defines the units of the magnetic flux $\Phi$ per
unit cell of the periodic potential, $\alpha =\Phi/\Phi _{0}$, where $\Phi
_{0}$ is the magnetic flux quantum. Hence, $\alpha$ also describes the
magnetic field if the periodic potential is fixed. Moreover, the size of the
sample is related to the period and $\alpha: N _{\phi }=L _{x}L
_{y}/\left(\alpha a_{0}^{2}\right),$ where $L _{x}$ and $L _{y}$ are the
length of the sample in the $x$ and $y$ directions. In what follows, we
would like to study the energy gaps in different magnetic fields or $\alpha$
for a fixed filling factor $\nu$. However, the size of the sample is not
fixed ($L _{x}$ and $L _{y}$ are not constant) when we study the system for
a continuous $\alpha$. 

\emph{Crystal phase in an infinite sample:} If we calculate the energy
spectrum for continuous $\alpha$ in a fixed-size sample, the energy gaps may
not be reliable, since both the energy spectrum and the gaps may be
size-dependent. To calculate the energy spectrum in an infinite sample, we
could work in the momentum space. To be consistent with the finite-size
study, we consider a crystal phase of the electron gas with the same
geometry as the periodic potential. The lattice constant of this electron
crystal $a$ is given by $a/\ell =\sqrt{2\pi n _{c}/\nu},$ where $n _{c}$ is
the electron number per crystal site. If the lattice constant of the
electron crystal is identical to the period of the potential $a=a _{0}$,
then the electron number per site is given by $n _{c}=\nu/\alpha.$

The Hamiltonian of the 2DEG in the Hartree-Fock approximation is written 
\begin{eqnarray}
H &=&\sum_{\sigma }E_{\sigma }\rho _{\sigma ,\sigma }\left( \mathbf{q}
=0\right) +\sum_{\sigma ,\mathbf{q}}V_{\rm ext}\left( \mathbf{q}\right) \rho
_{\sigma,\sigma }\left( \mathbf{q}\right)   \label{Hamiltonian} \\
&&+\overline{\sum_{\mathbf{q}}}\sum_{\sigma ,\sigma ^{\prime }}U_{H}\left( 
\mathbf{q}\right) \left\langle \rho _{\sigma ,\sigma }\left( -\mathbf{q}
\right) \right\rangle \rho _{\sigma ^{\prime },\sigma ^{\prime }}\left( 
\mathbf{q}\right)   \notag \\
&&-\sum_{\mathbf{q}}\sum_{\sigma ,\sigma ^{\prime }}U_{X}\left( \mathbf{q}
\right) \left\langle \rho _{\sigma ,\sigma ^{\prime }}\left( -\mathbf{q}
\right) \right\rangle \rho _{\sigma ^{\prime },\sigma }\left( \mathbf{q}
\right).  \notag
\end{eqnarray}
The density matrix operator is 
\begin{equation}
\rho _{\sigma ,\sigma ^{\prime }}\left( \mathbf{q}\right) =\frac{1}{N_{\phi }
}\sum_{X,X^{\prime }}e^{-\frac{i}{2}q_{x}\left( X+X^{\prime }\right) }\delta
_{X,X^{\prime }+q_{y}\ell ^{2}}c_{\sigma ,X}^{\dagger }c_{\sigma ^{\prime
},X^{\prime }},
\end{equation}
where operators $c_{\sigma ,X},c_{\sigma ,X}^{\dag }$ are the annihilation
and creation operators of electrons in valley-spin $\sigma $ and the guiding
center $X$. The Hartree and Fock interaction functions, $U_{H}$ and $U_{X},$
are defined by 
\begin{eqnarray}
U_{H}\left( \mathbf{q}\right)  &=&\frac{e^{2}}{\kappa \ell }\frac{1}{q\ell }
\left[ M_{q_{y}\ell ^{2}/2,-q_{y}\ell ^{2}/2}\left( q\right) \right] ^{2}, \\
U_{X}\left( \mathbf{q}\right)  &=&\frac{e^{2}}{\kappa \ell }\int dp\left[ M_{
\frac{p_{y}}{2},-\frac{p_{y}}{2}}\left( p/\ell \right) \right]
^{2}J_{0}\left( pq\ell \right) ,
\end{eqnarray}
where $J_{0}$ is the Bessel function. The external potential in such an
electron crystal is 
\begin{equation}
V_{\rm ext}\left( \mathbf{q}\right) =\frac{V_{0}}{2} M_{q_{y}\ell
^{2}/2,-q_{y}\ell ^{2}/2}\left( q\right).
\end{equation}

We define the Green's function $G_{\sigma ,\sigma ^{\prime }}\left(
X,X^{\prime },\tau \right) =-\left\langle Tc_{\sigma ,X}\left( \tau \right)
c_{\sigma ^{\prime },X^{\prime }}^{\dagger }\left( 0\right) \right\rangle ,$
where $T$ is the time order operator. Then at zero temperature, 
\begin{equation}
G_{\sigma ,\sigma ^{\prime }}\left( \mathbf{q,}\tau =0^{-}\right)
=\left\langle \rho _{\sigma ^{\prime },\sigma }\left( \mathbf{q}\right)
\right\rangle ,  \label{grho}
\end{equation}
where $G_{\sigma ,\sigma ^{\prime }}\left( \mathbf{q,}\tau \right) $ is the
Fourier transform of $G_{\sigma ,\sigma ^{\prime }}\left( X,X^{\prime },\tau
\right) $. In the Matsubara frequency\ $\omega _{n},$ the equation of motion
of the Green's function 
\begin{eqnarray}
\label{eqmotion}
&&\left( i\hbar \omega _{n}-E_{d}\right) G_{d,e}\left( \mathbf{q,}\omega
_{n}\right) =\hbar \delta _{d,e}\delta _{\mathbf{q},0}  \label{eom} \\
&&+\sum_{\mathbf{q}^{\prime }}V_{\rm ext}(\mathbf{q}^{\prime })e^{i\mathbf{q}
^{\prime }\mathbf{\times q}\ell ^{2}/2}G_{d,e}\left( \mathbf{q+\mathbf{q}
^{\prime },}\omega _{n}\right)   \notag \\
&&+\sum_{\sigma ,\mathbf{q}^{\prime }}U_{H}\left( \mathbf{q}^{\prime
}\right) \left\langle \rho _{\sigma ,\sigma }\left( -\mathbf{q}^{\prime
}\right) \right\rangle e^{i\mathbf{q}^{\prime }\mathbf{\times q}\ell
^{2}/2}G_{d,e}\left( \mathbf{q+\mathbf{q}^{\prime },}\omega _{n}\right)  
\notag \\
&&-\sum_{\sigma ,\mathbf{q}^{\prime }}U_{X}(\mathbf{q}^{\prime
})\left\langle \rho _{\sigma ,d}\left( -\mathbf{q}^{\prime }\right)
\right\rangle e^{i\mathbf{q}^{\prime }\mathbf{\times q}\ell ^{2}/2}G_{\sigma
,e}\left( \mathbf{q+\mathbf{q}^{\prime },}\omega _{n}\right), \notag
\end{eqnarray}
where $d,e$ are valley-spin indices, can be solved self-consistently (see 
\cite{cote2} for details). Then all the elements of the density matrix can 
be obtained according to Eq.~(\ref{grho}).

The density matrix completely describes the system with the Hamiltonian in
Eq.~(\ref{Hamiltonian}). The energy spectrum is thus obtained by solving the
equation of motion in Eq.~(\ref{eom}). When we estimate the energy gap, we
need to subtract the energy of the lowest unoccupied state by the energy of
the highest occupied state in the density of states (DOS). The relation
between the DOS $g$ and the retarded Green's function is \cite{cote3} 
\begin{equation}
g\left(\omega\right) = -\frac{N _{\phi}}{\pi}\sum _{\sigma}{\Im}\left[
G_{\sigma,\sigma}^{R}\left( \mathbf{0},i\omega _{n}\rightarrow \omega
+i0^{+}\right)\right].
\end{equation}

The crystallized electron gas was studied in monolayer graphene \cite{cote}
and graphene bilayer \cite{cote3} without the external potential. Wigner
crystals and skyrmion crystals can be found in those systems at non-integer
filling factors. Moreover, the skyrmion crystals were found in bilayer 
\cite{cote4} and trilayer graphene \cite{yafis} even for integer filling factors,
due to the Dzyaloshinskii-Moriya (DM) interaction. We employ the same method
to study the electron gas in the presence of the external potential. For a
strong potential the 2DEG may be crystallized with the same geometry
as the potential for integer filling factors even without the DM interaction.

In a recent experiment \cite{yu}, there is about 1$^{\circ}$ misalignment
between graphene and the BN substrate with the dielectric constant $\kappa=8$.
The period of the moir\'{e} pattern is about 100 times larger than the
lattice constant of graphene. Here we fix the period of the potential, $a
_{0}=30$ nm, and the amplitude of the potential, $V _{0}=20$ meV. We could
then neglect the Landau level mixing, since it is very weak. The period is
large enough to avoid the valley mixing. Hence, we could set the valley same
as spin. The valley pseudo-spin is conserved in the HFA. In order to be
consistent with the experimental results, we consider $\alpha \in \left[1,2
\right]$, so $1/\alpha \in \left[0.5,1 \right]$ ($\alpha$ in Ref.~\cite{yu}
is $1/\alpha$ in this paper). Because of the fractal pattern, the
noninteracting energy spectrum in the region $\alpha \in \left[ 1,2\right]$
is similar to that in the region $\alpha \in \left[0,1\right]$.

\textit{Finite-size study for} $\nu =-1$: For the finite-size study of the
energy spectrum in the HFA, we fix the size of the sample, and change the
magnetic field (or the parameter $\alpha$). The size of the sample is $6a
_{0}\times 6a _{0}$. For the filling factor $\nu =-1$ in the Landau level $
N=0$, there is $4N _{\phi}$ degeneracy if we do not consider the Zeeman
coupling, but there are only $N _{\phi}$ electrons in this LL, $N _{s}=N
_{\phi}.$ So in this sample, $N _{\phi}=\left[ 36/\alpha \right] $, where $
\left[ x \right]$ is the largest integer not exceeding $x$.

\begin{figure}[tbp]
\includegraphics[width=7cm]{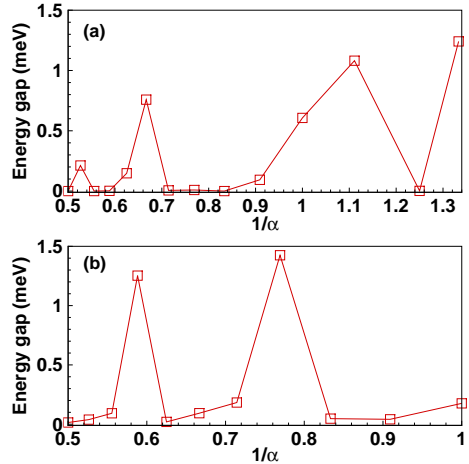}
\caption{The energy gaps for the filling factor $\protect\nu =-1$ with
different $1/\protect\alpha $. (a) The size of the sample is $6a _{0}\times
6a _{0}$. (b) An infinite sample.}
\label{nu-1}
\end{figure}

In Fig.~\ref{nu-1} (a), we show the energy gaps with different $\alpha$. 
The energy gap oscillates as the magnetic field increases. When $\alpha =1.5$
, which is equivalent to $\alpha=0.5$, a previous study \cite{vadym} showed
that the gap between the two bands would be open when the spin is polarized
and one valley is half-filled in the Hartree approximation. In this work, we
take the spin into consideration. The ground state is no longer spin
polarized. The Zeeman coupling is very weak (about $1$ meV) while the
amplitude of the external potential is $20$ meV. The potential is strong
enough to mix different spins. The gap for $\alpha =1.5$ is very small in
Fig.~\ref{nu-1} (a). This is because, intuitively, the two spins are mixed,
and the corresponding four bands (two for each spin) in one valley are
overlapped to close the Hartree gap. This mixed spin ground state will be
discussed in detail in the infinite-size study below.

\begin{figure}[tbp]
\includegraphics[width=7cm]{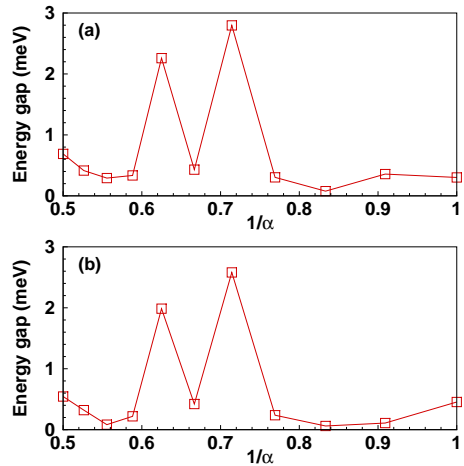}
\caption{Energy gaps (a) for $\protect\nu =0$ and (b) for $\protect\nu =4$.}
\label{nu0_4}
\end{figure}

\textit{Energy gaps in an infinite sample for} $\nu =-1$: The finite-size
study however, may not be very reliable. In fact, solution of Eq.~(\ref{eqmotion}) 
is predicated on the size of the sample being infinite. In Fig.~\ref{nu-1} (b), the 
energy gap also oscillates with $1/\alpha $. However, the amplitude and the peaks of
the oscillation are changed a little. This might be because the system is
size-dependent when the size is finite. For $\alpha =1.5$, the gap is also
very small, which is simialr to that of the fintie-size calculation.
Generally, the results of the two different calculations are similar.
Experimental results (Fig.~4 (c) in \cite{yu}) also show oscillations in the
energy gap, but the measured gap is nonzero for $\alpha =2$.

\textit{Energy gaps in an infinite sample for} $\nu =0,4$: We calculate for
filling factors $\nu =0,4$, in the LLs $N=0,1$, respectively. In these
cases, each LL is half-filled, i.e., there are $N_{s}=2N_{\phi }$ electrons
in the LL $0$ or $1$. The spinless picture is obviously not satisfied in
this case. We need to consider all the spins and valleys. Figure~\ref{nu0_4}
(a) and Fig.~\ref{nu0_4} (b) show the energy gaps for the filling factors $
\nu =0,4,$ respectively. These two curves are similar: the gaps are very
small except at two points $\alpha =1.4,1.6$. Note that for $\alpha =1.5$,
the energy gap is small, due to the same reason as what we explained for $
\nu =-1$. For the filling factor $\nu =0$, the numerical results are
different from the experimental results \cite{yu}, where the energy gap
curve looks like the energy cruve of the charged skyrmion excitation 
\cite{fertig}. However, in our calculations we can not obtain such a
skyrmion crystal ground state. It might be because the electron density is
much higher than what the skyrmion crystal was found numerically \cite{cote}.
The spin or pseudo-spin textures are suppressed by the high density
electron gas: (pseudo-) spin flipping can not decrease enough energy to
create a (pseudo-) spin texture. The reason why our numerical results differ
from the experiment is because perhaps the ground state in the $N=0$ LL is
not spin polarized without the external potential \cite{young}, and we only
consider a spin polarized liquid ground state here.

\begin{figure}[tbp]
\includegraphics[width=7cm]{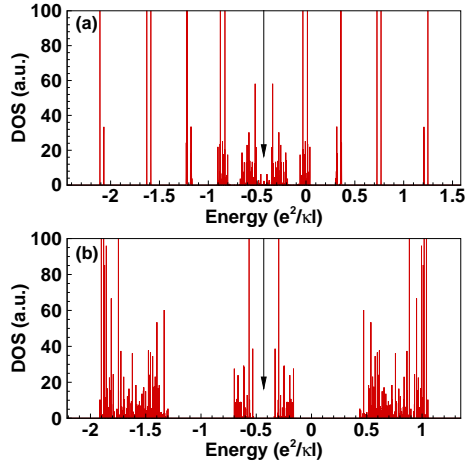}
\caption{(Color online) The DOS for $\protect\nu =4,$ and (a) for $\protect
\alpha =1.1$ and (b) for $\protect\alpha =1.4$. The arrows indicate the
Fermi level locations.}
\label{dos_nu4}
\end{figure}

For $\nu=4,$ our numerical results are similar to those observed in the
experiment \cite{yu}. In the DOS, we clearly see the energy band structures.
The DOS for $\overline{\alpha }\in \left[ 1.5,2\right]$ is similar to the
DOS for $\alpha =3-\overline{\alpha}$, so that we neglect the DOS for $
\alpha >1.5$. For simiplicity, we show the DOS curves for $\alpha =1.1$ and $
\alpha =1.4$ in Figs.~\ref{dos_nu4} (a) and \ref{dos_nu4} (b), respectively.
For $\alpha =1.1,$ there are ten bands, but the middle two bands touch at
the Fermi level. The energy gap is almost zero. Note that some bands far
away from the Fermi level split into two sub-bands, due to the Zeeman
coupling. For $\alpha =1.4,$ there is a gap between the middle two bands and
all bands mix both spins, opening a gap (about 2.5 meV).

\begin{figure}[tbp]
\includegraphics[width=8.0cm]{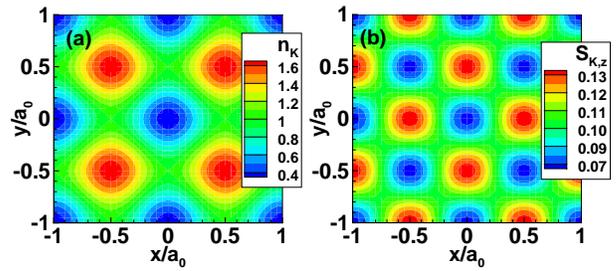}
\caption{(Color online) (a) Density profile for $\protect\alpha =1.4,\protect
\nu =4$, in the $K$ valley. The density is in units of $1/(2\protect\pi 
\ell^{2})$. (b) The spin field, in units of $\hbar /\left( 2\protect\pi 
\ell^{2} \right),$ in the $z$ direction $S _{K,z}\,\ $in the $K$ valley for $
\protect\alpha =1.4, \protect\nu =4$ are showed.}
\label{density_alpha1.4_nu4}
\end{figure}

We now define the spin field \cite{ezawa} in valley $\eta $ ($K$ or $
K^{\prime }$) 
\begin{eqnarray}
&&S_{\eta,x}+iS_{\eta,y}=\left\langle \rho _{(\eta,\uparrow ,(\eta,\downarrow )}
\left(\mathbf{r}\right)\right\rangle, \\
&&S_{\eta,z}=\left\langle \rho_{(\eta,\uparrow),(\eta,\uparrow )}\left( 
\mathbf{r}\right) -\rho_{(\eta,\downarrow),(\eta,\downarrow)}\left( 
\mathbf{r}\right) \right\rangle .
\end{eqnarray}
The density in valley $k$ is given by $n_{k}\left( \mathbf{r}\right)
=\sum_{s}\left\langle \rho_{(\eta,s),(\eta,s)}\left(\mathbf{r}\right)
\right\rangle,$ where $s$ is the spin index. The two valleys are completely
equivalent in our numerical results, so only the order parameters in the $K$
valley are shown in Fig.~\ref{density_alpha1.4_nu4}. The density profiles
have the same geometry as the external potential. There is no valley
coherence which is not showed in Fig.~\ref{density_alpha1.4_nu4}, i.e. $
\left\langle\rho_{\left(K,s\right),\left(K^{\prime},s^{\prime}\right)
}\right\rangle=0.$ The spin field contains no texture at all, $S_{\eta,x}=
S_{\eta,y}=0$, so the electron crystal is not a skyrmion crystal. Only
the $z$-components are nonzero, $S_{\eta,z}\neq 0$, and $S_{\eta,z}$ is
also crystallized.

Note that the maximum points of $S_{\eta,z}$ do not match the maximum
points of the density, where the minimum points of the external potential
are. At these points the potential decreases the kinetic energy for both
spins. The electrons with both spins overcome the repulsive interaction to
be localized by the potential. The density of electrons is minimum at the
sites where the energy of the potential is maximum. At the points where the
density of electrons is maximum or minimum, the spin field $S_{\eta,z}$ is
minimum [the blue dots in Fig.~\ref{density_alpha1.4_nu4} (b)].

In conclusion, we have studied the interacting Hofstadter's butterfly states
in a finite-size system in the HFA in order to study the spin/valley systems
such as graphene. We also used a method where the sample is infinite. 
The energy gaps in the finite-size study agree with the results of the electron
crystal qualitatively for filling factor $\nu =-1$. The excitation gap
oscillates with the increase of the magnetic field (or $1/\alpha $), similar
to a recent experimental observation. For half fillings, we employ the
crystal method to calculate the DOS of the system. In the $n=1$ LL, the
energy gap in a magnetic field (or $1/\alpha$) agrees well with the
experimental results. The osillation of the energy gap agrees qualitatively
with that in the experiment, which the nonintearcing picture without spin or
valley can not explain. Finally, we show the ground state of the electron
gas in the Hofstadter's butterfly state. The two spins are mixed while there
is no valley coherence in the system. This crystal phase is neither like a
Wigner crystal nor a skyrmion crystal. The electron gas tends to be in a
liquid phase, but the strong external potential crystallizes it. We propose
that this method is able to study the interacting (in the HFA) Hofstadter's
butterfly states with different periodic potentials in an infinite system
conveniently and efficiently, not only in graphene, but also in other Dirac
materials.

The work has been supported by the Canada Research Chairs Program of the
Government of Canada.

\end{document}